%% file: main.tex
\documentclass[manuscript, screen]{jair}
\setcopyright{cc}
\copyrightyear{2025}
\acmYear{2025}
\acmDOI{10.1613/jair.1.xxxxx}

\JAIRAE{Insert JAIR AE Name}
\JAIRTrack{Insert JAIR Track Name Here}
\acmVolume{4}
\acmArticle{111}
\acmMonth{8}
\received{31 May 2025}
\received[accepted]{18 August 2026}

\input{preamble}

\begin{document}
\title{Revisiting Conflict-Based Search with Continuous-Time}

\author{Andy Li}
\email{andy.li@monash.edu}
\orcid{0009-0000-1990-928X}
\monash

\author{Zhe Chen}
\email{zhe.chen@monash.edu}
\orcid{0000-0002-0425-3131}
\monash

\author{Mor Vered}
\email{mor.vered@monash.edu}
\orcid{0000-0001-5286-6509}
\monash

\author{Daniel Harabor}
\email{Daniel.Harabor@monash.edu}
\orcid{0000-0001-6828-7712}
\monash

\begin{abstract}
Multi-Agent Path Finding in Continuous Time (\mapfr) extends the classical MAPF
problem by allowing agents to operate in continuous time.
Conflict-Based Search with Continuous Time (CCBS) is a foundational algorithm for solving \mapfr optimally.
In this paper, we revisit the theoretical claims of CCBS and show the algorithm is incomplete, due to an uncountably infinite state space created by continuous wait durations.
We then prove that the constraint generation method employed by CCBS inherently leads to incompleteness
(i.e., there is no simple fix).
For further insight into these issues, we consider a related sub-problem, \mapfrdt, which
restricts wait durations to fixed amounts, and show that it is optimally solvable, including by CCBS.
It remains an open question whether \mapfr, which allows arbitrary wait durations, is optimally solvable.
\end{abstract}

\maketitle
\renewcommand{\shortauthors}{Li, Chen, Vered \& Harabor}
\input{body}

\printbibliography
\appendix
\input{sections/Appendix}
\end{document}

%% file: body.tex
\input{sections/intro}
\input{sections/Background}
\input{sections/problem}
\input{sections/imp}
\input{sections/discussion}

\input{sections/MPAF2MAPFR}
\input{sections/conclusion}

%% file: sections/intro.tex
\section{Introduction}

Multi-Agent Path Finding (MAPF) is the problem of planning paths for a set of
moving agents from initial states to goals while avoiding collisions.
MAPF is gaining increasing attention because of its significance in a wide range of application settings, both in the physical world such as warehouse management~\cite{kiva} and robotics~\cite{motionPlanning2018}, as well as in virtual environments such as computer games~\cite{harabor2022benchmarks}.
Classical MAPF models~\cite{stern2019multi} reduce these problems to a simplified (unit-cost (di)graph, grid) combinatorial core, which is nonetheless NP-hard~\cite{yu2013np_hard,banfi2017intractability,Nebel_nphardDigraph_2020}.

Conflict-Based Search (CBS) is currently the most widely used approach for optimally solving Classical MAPF ~\cite{cbs,icbs,LI2021Pairwise,cbs-beyond}.
To tackle the problem, CBS uses a two-level approach. At the high level, CBS searches over a Conflict Tree (CT), where each node represents a set of constraints on the actions of the agents. At the low level, it begins with each agent's individual shortest path as a lower bound and resolves conflicts by iteratively identifying colliding agent pairs and imposing a pair of sound constraints to eliminate collisions.
Unfortunately, classical MAPF plans are seldom applicable in practice as many physical applications operate in continuous environments, where motions have non-unit duration and their executions are not synchronised.

\mapfr extends the classical MAPF model by generalising the time dimension to real-valued domains, allowing agents to move in \textit{continuous} time and wait for any arbitrary amount of real-valued time duration.
Continuous Time CBS (CCBS)~\cite{ccbs} is claimed to be the first optimal algorithm for solving this problem.
The algorithm is widely popular in the literature and
several subsequent works provide further improvements \cite{improve-ccbs,Walker2020byclique,Walker2021cbics,walker2024byclique} and enhancements \cite{Yakovlev2024AACCBS,Kulhan2023quadchop}.
In this work, we show that the original constraint definition in CCBS~\cite{ccbs} leads to two distinct interpretations regarding wait motion: the first interpretation assumes wait durations are specified a priori, the second assumes durations are computed dynamically.
We demonstrate that, under \textit{both} of these interpretations, CCBS is either unsound, incomplete, or both.
The implications of our findings affect numerous recent works,
all of which implicitly rely on the correctness of the CCBS algorithm.
It is an open question whether there exists an alternative constraint-based reasoning model which can resolve these issues.

To place these findings in context, we close the paper with a broader discussion of the \mapfr problem.
We first distinguish the original \mapfr model, which permits arbitrary real-valued wait durations,
from the restricted \mapfrdt model, in which every wait motion has a fixed unit duration, a distinction that wasn't explicit in the literature.
We show that the search space of \mapfrdt remains countable and that the problem is optimally solvable, including by CCBS;
the incompleteness we identify is therefore specific to the continuous wait durations of the general \mapfr setting.
We then ask whether \mapfr is solvable at all.
Classical results in this area all depend on a simplified discrete action model \cite{kornhauser1984pebble,Yu2015pebblewithRotation,surynek2009bibox,wang2011mapp}.
However, there currently exists no known algorithm for \mapfr which is truly sound and complete.
We identify a number of important modelling differences, such as waiting agents that overlap neighbouring vertices and edges, and agents buffering along a shared edge, that prevent us from applying existing solvability arguments in a continuous time domain.
Finally, we briefly discuss the numerical issues that hinder the implementation and reproducible evaluation of \mapfr solvers in practice.

%% file: sections/Background.tex
\section{\mapfr Problem Definition}
We use the continuous-time MAPF problem definition defined by \citet{ccbs},  which takes a weighted graph $G=\{V,E\}$ and a set of $k$ agents $\{a_1, ...,a_k\}$ as input.
Every vertex $v$ in $G$, maps to a coordinate in a metric space $\mathcal{M}$, and every edge connects between two vertices.
Each agent has a unique start, $s_i \in V$ and goal vertex $g_i\in V$ with no two agents sharing the same start or goal vertices: $\forall i \in\{1,...,k\} \nexists j \in\{1,...,k\} \setminus i: s_i=s_j \vee g_i=g_j$.
An agent $a_i$ can start to execute a motion $m_i^j$ at $\tau_i^j$, i.e. $m_i^j@\tau_i^j$.
A plan $\pi_i$ of an agent $a_i$ is a sequence of motions $m_i^j$ with start times $\tau_i^j$, such that $\{ m_i^1@\tau_i^1,..., m_i^l@\tau_i^l\}$.
Each motion $m$ consists of $\langle m.\varphi,m.D \rangle$:
\begin{itemize}
\item $m.D$, a positive real number that indicates the duration of the motion.
\item $m.{\varphi}$, a motion function $m_{\varphi} : [0, m.D] \rightarrow \mathcal{M}$ that maps time to a coordinate in the metric space.
\end{itemize}

The motion function $m.{\varphi}(t)$ enables the agent to move at a non-constant speed and follow an arbitrary geometric curve.
However, in this paper, following the original CCBS assumption, all agents move in a straight line at a constant speed of 1 without any kino-dynamics.
Therefore, we will use $d(v,v')$ to denote the length of the edge $(v,v')$, which also indicates the cost of the moving motion along this edge.
Conveniently, the motion function is a linear function that outputs the coordinate of the agent by executing motion $m$ for duration $t$.

In general, a motion function comprises of either a moving motion or a waiting motion.
For a moving motion, $m.{\varphi}(0)$ returns the current vertex $v$ and $m.{\varphi}(m.D)$ returns the next vertex $v'$.
A waiting motion happens when $v'=v$ and $m.{\varphi}(t)$ returns the coordinate of $v$ for the entirety of $[0, m.D]$.
In the \mapfr problem, we assume a finite set of move motions, corresponding to the finite number of edges in $G$, with the motion duration being fixed and represented by the edge weights.
However, the set of wait motions is assumed to be uncountably infinite, as agents can wait for any positive real-valued amount of time, leading to an infinite range of possible wait durations.
In a valid plan $\pi_i$, the origin vertex $v^n$ of $m_i^n:(v^n,v'^{n})$ must coincide with the target vertex $v'^{n-1}$ of the previous motion $m_i^{n-1}:(v^{n-1},v'^{n-1})$.
Similarly, the starting time $\tau_i^n$ of motion $m_i^n$ is always equal to the finish time $\tau_i^{n-1} + m_i^{n-1}.D$ of the previous motion $m_i^{n-1}$, where $m_i^{n-1}.D$ denotes the duration of the previous motion.

A feasible solution to a \mapfr problem consists of a set of plans $\Pi = \{\pi_1, ..., \pi_k\}$ where each planned motion of every agent $a_i$ is conflict-free with respect to every planned motion of every other agent $a_j \neq a_i$.
A \textbf{conflict} $\langle a_i,a_j, m_i@\tau_i, m_j@\tau_j\rangle$ between two agents $a_i$ and $a_j$ occurs when a collision detection function $InCollision(m_i@\tau_i,m_j@\tau_j)$ returns $true$.
This indicates that there exists a time $t$ at which the locations of the two agents overlap.

Following the original CCBS assumption that all agents are circular with radius $r$, the collision checking function $InCollision(m_i@\tau_i,m_j@\tau_j)$ returns true if there exists a time $t$ such that the distance between the centres of the two agents $a_i$ and $a_j$, is smaller than $2r$, $||m_i-m_j||_2<2r$.
Note that if $InCollision(m_i@\tau_i,m_j@\tau_j) = true$, then there must exist two time points $t^c_s$ and $t^c_e$,  where $t^c_s$ is the first time when the distance between the two agents is exactly  $2r$, and $t^c_e$ is the second time when their distance is $2r$, or when one of the motions finishes.
We refer to the interval $I^c=(t^c_s,t^c_e)$ as the \textbf{collision interval}.
The objective is to compute a collision-free plan that minimises the \emph{Sum of all agents' Cost} (SoC),
where $SoC = \sum_{i\in [1,k]} et_i$ with $et_i$ indicating the end time of the last motion in path $\pi_i$.

\section{Background}
\subsection{CBS with Continuous Time}

CCBS is an extension of the original optimal MAPF solver CBS, specifically designed to solve the \mapfr problem optimally.
CCBS is a two-level tree search algorithm, where the high-level search is a best-first search on a binary \emph{Constraint Tree} (CT).
Each CT node $N=\{N.constraints,N.\Pi,N.g$\} contains:
\begin{itemize}
\item  $N.constraints$, a set of conflict-resolving constraints,
\item $N.\Pi$, a solution that satisfies these constraints, and
\item $N.g$, the sum of the individual costs (SoC) of $N.\Pi$.
\end{itemize}

At the low level of CCBS, the algorithm finds a temporal shortest path that satisfies all the constraints for each agent at the current node.
CCBS employs \emph{Safe Interval Path Planning} (SIPP) \cite{sipp} to compute these paths.
The process begins by generating a root CT node with no constraints ($N.constraints=\phi$) and a solution consisting of individual shortest plans for all agents.
It then iteratively selects a node $N$ with the smallest $N.g$.
If $N$ contains no conflicts, it is deemed the goal node.
Otherwise, a conflict from  $N.\Pi$, such as $\langle a_i,a_j, m_i@\tau_i, m_j@\tau_j\rangle$, is selected for resolution.

CCBS then generates two child CT nodes $N_i$ and $N_j$, each appending a new conflict-resolving constraint $C$ based on $N.constraints$, so that:
\begin{itemize}
\item $N_i.constraints = N.constraints \cup C_i$
\item $N_j.constraints = N.constraints \cup C_j$
\end{itemize}

The paper proposes a pair of \textbf{Motion Constraints} to resolve the conflict:
\begin{itemize}
    \item $C_i = \overline{\langle a_i, m_i, I_i \rangle}$
    \item $C_j = \overline{\langle a_j, m_j, I_j \rangle}$.
\end{itemize}
    The unsafe interval $I_i:[t_{is},t_{ie})$ (resp. $I_j:[t_{js},t_{je})$) for agent $a_i$ (resp. $a_j$) is the constraint interval for $C_i$ (resp. $C_j$).
    If the starting time $\tau_i$ (resp. $\tau_j$) of motion $m_i$ (resp. $m_j$) falls within  $I_i$ (resp. $I_j$), it will result in a conflict with $m_j@\tau_j$ (resp. $m_i@\tau_i$).
    We will use $t_s$ and $t_e$ to denote the starting and end time for an interval $I$, i.e., $I=[t_s,t_e)$.
    In this case $t_{is}=\tau_i$, and $t_{js}=\tau_j$.
A motion constraint $C_i:\overline{\langle a_i, m_i, I_i \rangle}$ (resp. $C_j:\overline{\langle a_j, m_j, I_j \rangle}$) indicates that agent $a_i$ (resp. $a_j$) cannot start motion $m_i$ (resp. $m_j$) within the \emph{unsafe interval} $I_i$ (resp. $I_j$).

\subsection{The Claims of CCBS}

We begin with the following definition from CCBS~\cite{ccbs}:
\begin{definition}[Sound pair of constraints]
    \label{def:sound}
    Given a \mapfr problem, a constraint pair is sound iff in every feasible solution, at least one of these constraints hold.
\end{definition}

CCBS then makes the following claims:
\begin{claim}
\label{clm:sound}
Every pair of constraints used by CCBS for resolving a conflict is sound, and any \mapfr solution that violates both constraints must have a conflict:

\begin{lemma}
    \label{lemma:sound}
    For any CCBS conflict $\langle a_i,a_j, m_i@\tau_i, m_j@\tau_j\rangle$, and corresponding unsafe intervals $I_i$ and $I_j$, the pair of CCBS constraints $\overline{\langle a_i, m_i, I_i  \rangle}$ and $\overline{\langle a_j, m_j, I_j  \rangle}$ is sound.
\end{lemma}
\end{claim}

\begin{claim}
\label{clm:optimality}
CCBS is complete and is guaranteed to return an optimal solution if one exists.
\end{claim}

To support Claim~\ref{clm:optimality}, the original paper proved Lemma~\ref{lemma:condition}.
\begin{lemma}
    \label{lemma:condition}
For a CT node $N$, letting $\pi(N)$ be all valid \mapfr solutions that satisfy $N.constraints$,
$N.g$ be the cost of $N$,
and $N_1$ and $N_2$ be the children of $N$, the following two conditions hold for any $N$ that is not a goal node.
\begin{enumerate}
    \item $\pi(N)=\pi(N_1) \cup \pi(N_2)$ \label{cond:1}
    \item $N.g \leq min(N_1.g,N_2.g)$ \label{cond:2}
\end{enumerate}

\end{lemma}
The first condition states that no collision-free solution should be eliminated.
This holds because $N_1$ and $N_2$ are constrained by a sound pair of constraints (Definition \ref{def:sound} and Lemma \ref{lemma:sound}).
The second condition states that, during the search, node cost can only monotonically increase, it holds because $N.solution$ by construction is the lowest cost solution that satisfies the constraints in $N$, and the constraints in $N_1$ and $N_2$ are a superset of constraint in $N$.
The first condition ensures that any valid solution can be reached through one of the un-expanded CT nodes.
Since CCBS conducts a best-first search over the CT and prioritises expanding CT nodes with the lowest cost, the second condition enables CCBS to guarantee finding an optimal \mapfr solution.
Together, these two conditions aim to ensure the completeness of CCBS, meaning that if an optimal solution exists, it will be found.

%% file: sections/problem.tex
\subsection{Examining the Theoretical Proof Limitations  of CCBS}
It is important to note that there is ambiguity in the MAPF community regarding the description of the constraints used to resolve conflicts between a moving agent and a waiting agent.
This ambiguity gives rise to two different interpretations of the constraint applied to the waiting agent. For the sake of clarity, we assign distinct names to each interpretation:
\begin{itemize}
    \item \textit{Motion Constraint}: In this model, the respective constraint forbids the start of a motion with a specific duration (whether movement or wait) within the constraint interval. We will analyse CCBS with motion constraints in Section \ref{Section: Motion Constraint} with formal definitions.
    \item \textit{Vertex Range Constraint}: In this model, the respective constraint forbids any waiting motion from overlapping with the constraint interval, regardless of its duration or when it starts. We analyse CCBS with vertex range constraints in Section \ref{Section: Vertex Range Constraint} with formal definitions.
\end{itemize}
In the following two sections, we detail the differences between these two interpretations and formally demonstrate that they are either unsound, incomplete, or both.

\section{Interpretation 1: Motion Constraints}
\label{Section: Motion Constraint}
This interpretation originates from strictly adhering to the problem and conflict definitions as stated in the original CCBS paper.
Under this interpretation, wait and move motions are modelled in the same way, where each motion has a specific duration.

Suppose we are given a CT node $N$ with cost $N.g$ and a collision $c=\langle a_i, a_j, w_i@\tau_1, m_j@\tau_2 \rangle$ between a wait motion $w_{i}=\langle (v_1,v_1), w_i.D \rangle$ for $a_i$ and a move motion $m_{j}=\langle (v_2,v_3), ||(v_2,v_3)||_2 \rangle$ for $a_j$, where $||(v_2,v_3)||_2$ is the length of edge $(v_2,v_3)$.
CCBS then generates a sound pair of \emph{motion constraints}:

\begin{itemize}
    \item $\overline{\langle a_i, w_i, [\tau_1,t_{1e}) \rangle}$, which forbids $a_i$ to take the wait motion $w_i$, with duration $w_i.D$, from $\tau_1$ to $t_{1e}$, and
    \item $\overline{\langle a_j, m_i, [\tau_2,t_{2e}) \rangle}$, which forbids $a_j$ to take the move motion $m_j$ from $\tau_2$ to $t_{2e}$.
\end{itemize}
The intervals $[\tau_1, t_{1e})$ and $[\tau_2, t_{2e})$ are the maximal \emph{unsafe intervals} for $w_i$ and $m_j$, respectively.
If agent $a_i$ starts $w_i$ at any time within $[\tau_1, t_{1e})$, it will inevitably collide with agent $a_j$ starting $m_j$ at $\tau_2$.
Similarly, if agent $a_j$ starts $m_j$ at any time within $[\tau_2, t_{2e})$, it will inevitably collide with agent $a_i$ starting $w_i$ at $\tau_1$.

Although a sound pair of motion constraints ensures that no collision-free solutions are eliminated during splitting, this alone is not sufficient for CCBS to find an optimal solution.
In particular, recall that a motion is defined by both a motion function and its duration.
That means each real-valued waiting duration corresponds to a distinct wait motion.
Thus, each constraint only forbids a wait motion with a single duration at vertex $v_1$.
However, there are uncountably many wait motions at $v_1$ that can collide with the conflicting motion $m_j@\tau_2$
and it becomes infeasible to resolve such conflicts through motion constraints alone:

\begin{lemma}
    \label{lemma:inf expansion}
    Let $N'$ with a motion constraint $\overline{\langle a_i, w_i, [t_{is},t_{ie}) \rangle}$ be a child node of $N$ with $a_i$ and $a_j$ collide in the collision interval $(t^c_s,t^c_e)$.
    The constraint eliminates solutions where $w_i$ conflicts with $m_j@\tau_j$, but permits an infinite number of solutions that have an alternative
    wait motion $w_i' = \langle (v_i,v_i), w_i.D + \delta \rangle$ which conflicts with $m_j@\tau_j$, where $\delta \in (w_i.D-t^c_s + \tau_i,\infty) \setminus \{0\}$.
\end{lemma}

\begin{proof}
    Since the duration of wait motions can be any positive real number, there is an uncountable infinite number of choices on $\delta$.
    This leads to an infinite number of choices on $w_i'$ with $\delta$ in the given range, i.e. collision happens when $\tau_i + w_i.D - \delta > t^c_s$.
    For any wait motion $w_i'$ that executes at any time $\tau' \in [t_{is},t_{ie})$, it will collide with $m_j@\tau_n$, since the wait interval of $a_i$ waiting on $v_i$ is $[\tau', \tau'+w_i'.D)$, which always covers the duration of executing $w_i$ at the same time and $w_i$ collides with $m_j@\tau_2$.
\end{proof}

A main consequence of Lemma~\ref{lemma:inf expansion} is that CCBS must explore an infinite number of nodes and resolve an infinite number of conflicts, in order to advance the lower bound and ultimately identify an optimal solution, if one exists. This leads to non-termination, making the algorithm incomplete under these conditions.
More formally:

\begin{theorem}
\label{theorem:1}
    Given a CT rooted at $N$, if there exists an optimal solution node $N^*$ with cost $c^* = N.g + \Delta$ and $\Delta$ is a non-zero positive real number, CCBS, which uses pairs of \emph{motion constraints} to resolve conflicts, has to explore an infinite number of nodes to find $N^*$ if $N$ has conflicts between a pair of wait and move motion.
\end{theorem}

\begin{proof}
    CCBS searches in a best-first manner on solution cost $g$. To terminate the search and find an optimal
    solution node $N^{*}$, it must eliminate any collision solution with a cost smaller than $c^{*} = N.g+\Delta$ in its frontier.
    Assuming $a_i$ with wait motion $w_i@\tau_i$ is colliding with $a_j$ with move motion $m_j@\tau_j$,
    Lemma~\ref{lemma:inf expansion} shows that one child node of $N$ permits an infinite number of solutions, each with a $w_i' \neq w_i$ colliding with $m_j@\tau_j$.
    As long as $\Delta > 0$, there exists an infinite number of choices of $\delta \in [0, \Delta]$.
    Using a wait motion $w_i'=\langle w_i.\varphi, w_i.D + \delta\rangle$ to replace the $w_i$ in the solution of $N$, the resulting solution has cost increased by at most $\delta$.
    Therefore, we have an infinite number of collision solutions that have costs smaller than $N.g+\Delta$.
    As a pair of \emph{motion constraints} only removes collision solutions with one choice of wait motion duration, the algorithm requires an infinite number of expansions to remove all collision solutions with a cost smaller than $c^{*} = N.g+\Delta$ in its frontier.%
\end{proof}

To sum up, CCBS may fail to terminate and return a solution when it strictly resolves conflicts using \emph{motion constraints}.
While real-world computers operate over finite-precision representations rather than true real numbers,  CCBS would still need to explore every representable wait duration that could potentially resolve a conflict.
This is nearly impossible, as each expansion can exponentially increase the remaining search space, effectively doubling the workload with every new branch.

%% file: sections/imp.tex
\section{Interpretation 2: Vertex Range Constraint}
\label{Section: Vertex Range Constraint}

This interpretation is derived from Example \textit{4.1.3. SIPP for the low-level of CCBS}, which
appears in the CCBS paper~\cite{ccbs}. In this model wait motion have dynamically computed wait durations
and collisions with waiting agents are resolved by generating a corresponding {\em vertex range constraint}.
This approach is commonly adopted within the community, as evidenced by its use in many subsequent works \cite{walker2024byclique,combrink2025prioritizedplanningcontinuoustimelifelong,Yakovlev2024AACCBS}, including in the original implementation of CCBS\footnote{https://github.com/PathPlanning/Continuous-CBS}.
While vertex range constraints often lead to empirical termination, a closer
examination reveals that this model is also incomplete, although the problem arises in a
different manner than for motion constraints (which we discuss in Section~\ref{Section: Motion Constraint}).
Concretely, a vertex range constraint forbids the existence of an agent $a_i$ on a vertex $v$ within a given \emph{time interval} $I=(t_s,t_e)$.
We use $\overline{\langle a, v, I  \rangle}$ to denote such a constraint.
Whereas a motion constraint forbids the starting of a single wait motion (with a known duration), a vertex range constraint forbids any wait motion $\langle (v,v), D \rangle$ at any time $\tau$ as long as $[\tau,\tau+D]$ overlaps with $I$.

Consider the same CT node $N$ (appearing in Section~\ref{Section: Motion Constraint}) with the same conflict $c=\langle a_i, a_j, w_i@\tau_i, m_j@\tau_j \rangle$. Under this interpretation, CCBS will introduce a pair of \emph{vertex range and motion constraints}:
\begin{itemize}
    \item $\overline{\langle a_i, v_i, (t_s,t_e)  \rangle}$, a {vertex range constraint} which forbids agent $a_i$ from occupying $v_i$ during $(t_s,t_e)$, and
    \item $\overline{\langle a_j, m_j, [t_{js},t_{je}) \rangle}$, a {motion constraint} which forbids agent $a_j$ from executing motion $m_j$ during $[t_{js},t_{je})$.
\end{itemize}
In this constraint pair, $(t_s,t_e)$ is the collision interval between $w_i@t_i$ and $m_j@t_j$, and $[t_{js},t_{je})$ is the unsafe interval for $a_j$.
Executing $m_j$ at any time in this range must collide with $w_i@\tau_i$ for agent $a_i$.
A detailed example in Appendix \ref{App: collision interval} shows how CCBS computes
the collision interval, closely following the approach derived by \cite{walker2019collision}.
We also note that there is a subtle variation for the constraining time range used in the vertex range constraint.
According to the unsafe interval definition (Definition 3) and constraint notation (Lemma 1) from the original paper, the constraining interval always begins at the start time of the conflict move, i.e. $\overline{\langle a_i, v_i, [\tau_i,t_e)  \rangle}$.
However, the interval $[\tau_i, t_e)$ is strictly larger than $[t_s, t_e)$, which we will later show already excludes some valid solutions.
Therefore, using $[\tau_i, t_e)$ as the constraint interval would also eliminate feasible solutions.

\subsection{Elimination of Feasible Solutions}
\begin{figure}[h]
    \centering
    \begin{subfigure}[t]{0.33\columnwidth}
        \centering
        \includegraphics[width=1\columnwidth]{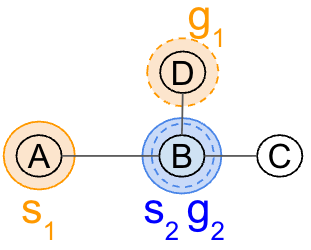}
        \caption{}
        \label{fig: wait exp space}
    \end{subfigure}%
    \hfill
    \begin{subfigure}[t]{0.25\columnwidth}
        \centering
        \includegraphics[width=0.9\columnwidth]{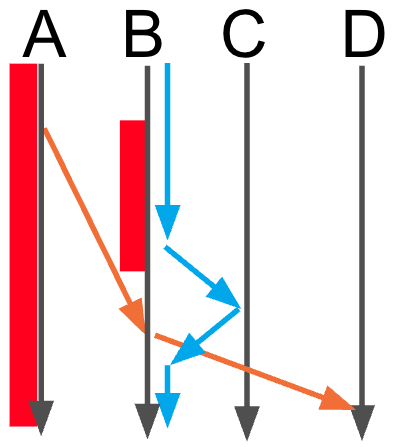}
        \caption{}
        \label{fig: wait exp broken time}
    \end{subfigure}
    \hfill
    \begin{subfigure}[t]{0.25\columnwidth}
        \centering
        \includegraphics[width=0.96\columnwidth]{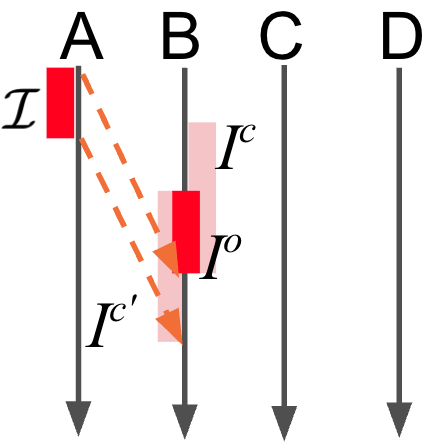}
        \caption{}
        \label{fig: wait exp sound time}
    \end{subfigure}
    \caption{(a) Agent $a_1$ performs motion $m_1= \langle (A,B),d_{AB} \rangle@0$, while agent $a_2$ has stopped at its goal $\langle (B,B), \infty \rangle@0$, causing a collision. (b) Motion and vertex range constraints with a pair of collision-free solutions. (c) \emph{Shifting constraints}. Constraints are represented by red regions and collision intervals by pink regions.
    }
    \label{fig:wait_exp_prob}
\end{figure}

We observe that the combination of a vertex range and motion constraints pair is not sound, as it eliminates feasible solutions.

An example illustrating the problem is shown in Figure \ref{fig:wait_exp_prob}.
Figure \ref{fig: wait exp space} shows the spatial graph, while Figure \ref{fig: wait exp broken time} illustrates the constraints and agents' movements over time, with time flowing from top to bottom.
In this example. agent $a_1$ needs to move from location $A$ to $B$ and then to $D$ while agent $a_2$ stops at its goal at location  $B$, causing a collision.
The existing implementation generates two constraints:
\begin{enumerate}
    \item
        The first constraint, $\overline{\langle a_1, m_1, [0,\infty)  \rangle}$ prevents agent $a_1$ from  moving to $B$ (shown as red region on $A$, Fig\ref{fig: wait exp broken time}).
    \item
        The second constraint, $\overline{\langle a_2, B, (t_2,t_3) \rangle}$, prevents agent $a_2$ from occupying $B$ within the time range of $(t_2,t_3)$ (shown as red region on $B$, Fig\ref{fig: wait exp broken time}).
\end{enumerate}
These constraints eliminate the collision-free solution where agent $a_1$ waits for a while and then departs (represented by the orange arrowed lines), and agent $a_2$ also waits before leaving and returning (blue arrowed lines).
This example empirically proves that a pair of \emph{vertex and motion constraints}, is not complete.
Such a solution may seem to be provably suboptimal for the problem when there are only two agents.
However, this is not the case for instances with more than two agents.
A valid solution may require both agents wait in order to provide the clearance for the third agent.

    \subsection{Counter Examples}
    \begin{figure}[h]
        \centering
        \includegraphics[width=0.5\columnwidth]{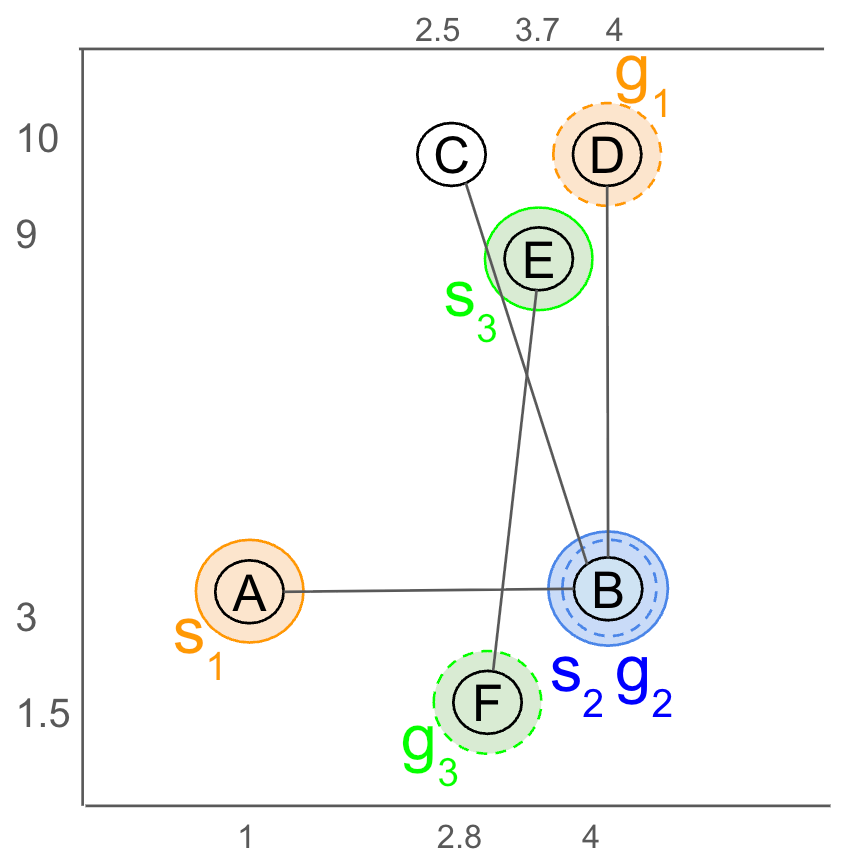}
        \caption{Agent $a_1$ (orange) is trying to move from $A$ to $D$, agent $a_2$ (blue) is trying to move from $B$ to $B$ (same starting and goal vertex) and agent $a_3$ (green) is trying to move from $E$ to $F$. The coordinates of vertices are: $A=(1,3)$, $B=(4,3)$, $C=(2.5,10)$, $D=(4,10)$, $E=(3.7,9)$ and $F=(2.825,1.5)$}
        \label{fig: counter example}
    \end{figure}
    Here, we will provide an example showing how eliminating such a situation could lead to eliminating all the feasible solutions, and yields CCBS incomplete.
    Fig \ref{fig: counter example} is modified based on the example shown in Fig \ref{fig: wait exp space} by moving vertex $C$ closer to $D$ and introducing another agent $a_3$ moving from $E$ to $F$. In this example, the agent on vertex $E$ will block the traversal on $(B,C)$ and $(B,D)$.

    If we were to run the CCBS algorithm on this instance, the first conflict would arise between agent $a_1$ and $a_2$ on $m_1=\langle (A,B),3 \rangle @ 0$, and $w_2\langle (B,B),\infty \rangle @ 0$.
    CCBS will generate $c_1=\overline{\langle a_1,m_1,[0,\infty) \rangle}$ for the moving agent, because the $a_2$ is waiting until $\infty$.
    This child node contains no feasible solution, as $a_1$ is not allowed to move at all.

    On the other branch, CCBS will generate $c_2=\overline{\langle a_2,B,[2,3) \rangle}$ for the waiting agent, because the two agents start to overlap at $t=2$, and the $m_1$ finishes at $t=3$.
    This constraint only permits the solution where $a_2$ departs from its $B$ to one of its neighbours before $t=2$.
    In this instance, $B$ has three neighbours $A$, $C$ and $D$.
    There is no solution for $a_2$ if it departs from $B$ to $A$, because $a_1$ is occupying $A$, and $A$ has no neighbours other than $B$.
    If $a_2$ moves to $C$ or $D$, it will collide with $a_3$.
    $a_3$ will clear the path of $(B,C)$ at $t=6.05$, thus if $a_2$ departs any time earlier than $6.05$ to $C$ or $D$, it will inevitably collide with $a_3$.
    Formally speaking, if $a_2$ replanned its path to move to $C$, i.e. $m_3=\langle (B,C),7.16 \rangle @ 0$, then CCBS will generate a corresponding constraint $c_3=\overline{\langle a_2,m_3,[0,6.05) \rangle}$.
     Similarly, $a_3$ will clear the path of $(B,D)$ at $t=6.09$, thus if $a_2$ replaned its path to move to $D$, i.e. $m_4=\langle (B,D),7 \rangle @ 0$, then CCBS will generate a corresponding constraint $c_4=\overline{\langle a_2,m_4,[0,6.09) \rangle}$.
    Therefore, either way $a_2$ has to wait for at least $6.05$.

    However, combining $c_2$ earlier, $a_2$ can only wait up to $t=2$, which means $a_2$ has no solution in this $c_2$ branch either.
    This results in the child node that contains $c_2$ to also containing no feasible solution.
    Therefore, the branching of $c_1$ and $c_2$ results in no feasible solution left in the search space.

    Note that CCBS will also generate respective constraints for $a_3$, when generating $c_3$ and $c_4$ for $a_2$; however, it's clear that putting $a_3$ wait at $E$ does not help for solving this problem instance, as it is still not possible for $a_2$ to move anywhere.
    Thus, we will not go through the full CCBS CT tree in this case.

    This counter example shows that vertex range constraint, $c_1$ and $c_2$, eliminates the possibility of both agents, $a_1$ and $a_2$, waiting, which is a necessary part of all feasible solutions for this example.

    Now we show a solution, which is sub-optimal due to the handcraft nature, but feasible solution, to this problem instance.
    For simplicity, we use the notation $v@\tau \rightarrow v'@\tau'$ to represent a motion $\langle m.D,(v,v') \rangle @\tau $ where $\tau' =\tau + m.D$.\\
    $a_1$:A@0 $\rightarrow$ A@9 $\rightarrow$ B@12 $\rightarrow$ D@19\\
    $a_2$:B@0 $\rightarrow$ B@8 $\rightarrow$ C@15.2 $\rightarrow$ C@19 $\rightarrow$ B@26.2\\
    $a_3$: E@0 $\rightarrow$ F@7.6 \\
    One can verify this solution is indeed collision-free by following the calculation in \cite{walker2019collision}.
    This feasible solution is eliminated by the $c_1$ and $c_2$ constraint pairs, as $a_1$ waited until $t=9$ and $a_2$ waited until $t=8$.

    Therefore, this counterexample proved that the vertex range constraint is not sound as it eliminates feasible solutions.
    Consequently, CCBS with the vertex range constraint is not complete, as it does not guarantee to find a solution if one exists.

\section{Shift Constraint}
In this section, we propose a method to construct a pair of {motion and vertex range constraints}, which will not remove collision-free solutions.
We call these a {\em sound pair of shifting constraints}.
The ability to safely eliminate more than one single in-collision solution at a time brings us a step closer to resolving the theoretical difficulties in CCBS.
    However, we also discovered some non-trivial properties of \emph{shifting constraints}, which makes proving the completeness of CCBS with shift constraints difficult.

    \subsection{Construction of Shift Constraints}
Given a move motion $m_j@\tau_j$, it collides with an agent with a waiting motion $w_i@\tau_i$ which occupies vertex $v$ within the \emph{collision interval} of $I^c=(I^c_s, I^c_e)$.
If we shift or delay the move motion $m_j$ to start at $\tau_j + \delta$,
it results a new \emph{collision interval} ${I^c}' = (t^c_s+\delta, t^c_e+\delta)$ with $w_i@\tau_i$, as shown in Figure~\ref{fig: wait exp sound time}.
We call $\mathcal{I}=[\tau_j, \tau_j+\delta)$ a \emph{shift interval}, and then define an \emph{overlapping interval} $I^o = (t^c_s+\delta, t^c_e) = I^c \cap {I^c}'$.
The \emph{overlapping interval} $I^o$ is a subset of or equal to a conflict interval $I^c$ between any motion $m_j@(\tau_j + d),  d \in [0,\delta]$ and $w_i@\tau_i$,
thus, any motion $m_j@(\tau_j + d),  d \in [0,\delta)$ conflicts with another agent exists on $v$ in range $I^o$.
We thus define a pair of \textbf{Shifting Constraints} between $m_j$ and any agent $a_i$ on $v$ as:
\begin{itemize}
    \item $\overline{\langle a_i, v, I^o \rangle}$, which forbids $a_i$ from waiting at $v$ in the interval $I^o$
    \item $\overline{\langle a_j, m_j, \mathcal{I} \rangle}$, which forbids $a_j$ from starting motion $m_j$ in the
        interval $\mathcal{I}$.
\end{itemize}
    Note that, by referring to Figure~\ref{fig: wait exp sound time} and the definition of \emph{shifting constraints}, it is obvious that the choice of the range of $\mathcal{I}$ decides the range of $I^o$, and the relation of their sizes is

    \begin{equation}
       \label{eq: interval size}
       |I^c| = |I^o| + |\mathcal{I}|
   \end{equation}

    Here we use $|I|$ to represent \textbf{the size of the interval} $|I|=|t_e-t_s|$, where $t_s$ and $t_e$ are the starting and ending times of the interval $I$.

    \subsection{Property of Shift Constraints}
    In this subsection, we prove the soundness of the shift constraint.
    However, as shown above, the soundness of the constraint alone is necessary but not sufficient for the completeness of CCBS.
    We therefore highlight the remaining limitation of the shift constraint, namely that it cannot fully constrain the collision interval.
\begin{theorem}
    \label{theorem:2}
    A pair of \emph{Shifting Constraints} is sound, as any motions violating both constraints must be in collision.
\end{theorem}
\begin{proof}
    Following the definition of \emph{Shifting Constraints}, given any start time $\hat{\tau_j} \in \mathcal{I}$ for motion $m_j$.
    The collision interval that $m_j@\hat{\tau_j}$ collides with agent $a_i$ staying at $v$ is $\hat{I^c} = I^c+x = (t^c_s+x, t^c_e+x)$, where $x=\hat{\tau_j}-\tau_j$ is the difference between the new and original move motion start time.
    Since $0 \leq x \leq \delta$, $I^o = (t^c_s+\delta, t^c_e)$,
    $t^c_s+x < t^c_s+\delta$ and $t^c_e+x > t^c_e$,
    thus $I^o$ is always a subset of \emph{collision interval} $\hat{I^c}$,
    indicating that any agent at the vertex $v$ in the range $I^o$ collides with any move motion $m_j@\hat{\tau_j}$.
\end{proof}

Next, we show that given a constraint interval on the motion constraint (equiv. vertex range constraint), if the range of the corresponding vertex range constraint (equiv. motion constraint) goes beyond the definition of a pair of \emph{shifting constraints}, collision-free solutions are eliminated.

\begin{property}
    \label{lemma:max_constraint_range}
    Given a motion $m_j$ (equiv. vertex $v$) of agent $a_j$ (equiv. $a_i$) constrained in an arbitrary $\mathcal{I}$ (equiv. $I^o$), and the corresponding vertex $v$ (equiv. motion $m_j$) for agent $a_i$ (equiv. $a_j$) is constrained in interval ${I^o}'$ (equiv. $\mathcal{I}'$). If there exists a time $\hat{t} \in {I^o}'$ (equiv. $\hat{t} \in \mathcal{I}'$), but $\hat{t} \notin {I^o}$ (equiv. $\hat{t} \notin \mathcal{I}$), meaning $\hat{t}$ does not exist in the constraint interval defined by \emph{shifting constraints}, collision-free solutions are eliminated.
\end{property}
\begin{proof}
    For the case of given motion $m_j$ constrained by $\mathcal{I}$, to have collision-free solutions be not eliminated, we must guarantee that $\hat{t}$ is not included in any collision interval resulting from the motion $m_j$ starting at any time $t_j' \in \mathcal{I}$.
    However, the definition of \emph{shifting constraint} shows that,
    $\hat{t} \notin {I^o}$ means that $\hat{t}$ is either earlier than the collision interval start time when move motion $m_j$ is started at a time close or equal to $\tau_j + \delta$, or is later than the collision interval end time when move motion $m_j$ is started at a time close or equal to $\tau_j$.
    Thus collision-free solution is eliminated by such a pair of constraints.
    The same logic applies when given the vertex range constraint interval $I_o$ on $v$ and there exists a time $\hat{t} \in \mathcal{I}'$ and  $\hat{t} \notin \mathcal{I}$.
\end{proof}

    \subsection{Termination Failure of Shift Constraints}
    We have demonstrated that \emph{shifting constraints} are a pair of sound constraints, and any pair of motion and vertex constraints beyond this definition will eliminate collision-free solutions.
    We now demonstrate that these properties may ultimately lead to termination failure, thus incompleteness for CCBS.

\begin{theorem}
    \label{thm:shifting-failure}
    Resolving conflicts between move motion and wait motion using sound pairs of \emph{shifting constraints} leads to termination failure.
\end{theorem}
\begin{proof}
    Following the Eq. \ref{eq: interval size} in the definition of \emph{shifting constraints}, the size collision interval is divided into two portions: the interval for motion constraint $I^o$, and the interval for vertex range constraint $\mathcal{I}$.
    \emph{Shifting constraints} will suffer different types of termination failure when $\mathcal{I}=0$ or $\mathcal{I}>0$:
    \begin{itemize}
        \item when $|\mathcal{I}| = 0$, the constraint interval $\mathcal{I}$ on $m_j$ becomes $[t,t)$ for some real-value $t$.
            This is mathematically incorrect, thus can be trivially interpreted as no constraint is introduced for the moving agent. This will result in an infinite loop, as no progress is made if no constraint is added. Alternatively, we can interpret $|\mathcal{I}| = 0$ as $[t,t]$, where the interval only contains one single real number (constraint on a single time).
            However, this constraint does not eliminate the conflict between any moving agent $a_j$ with motion $m_j@(t_j+d)~|~d \in (0,|I^c|] $ and the waiting agent $a_i$.
            As the amount of $d \in (0,|I^c|]$ is infinite, the approach suffers a similar infinite expansion problem to the one stated in Lemma~\ref{lemma:inf expansion}, leading to infinite expansions.
        \item when $|\mathcal{I}| > 0$, then the constraint interval $I^o$ on vertex $v$ cannot eliminate the collision solution that $a_j$ starts $m_j@\tau_j$ and $a_i$ occupies $v$ at $t^c_s$, as $t^c_s \notin I^o$.
            Therefore, the conflict is not fully resolved in the child node with the vertex range constraint; only a portion of the collision interval is eliminated.
            Fig \ref{fig: shift constraint example} shows an example of this behaviour.
            Thus, the expansion of this child node will eliminate another portion of the collision interval, as well as its child node, and so on.
            This results in Zeno's paradox, and the algorithm never terminates.
            Section \ref{Sec: example of shift fail} gives a working example of this behaviour.
    \end{itemize}
    Therefore, \emph{shift constraints} always have the conflict remaining in one of their child nodes and lead to infinite expansions.
\end{proof}

\subsection{Example Shift Constraint Failure \label{Sec: example of shift fail} }

\begin{figure}[t!]
    \centering
    \begin{subfigure}[t]{0.18\columnwidth}
        \centering
        \includegraphics[width=1\columnwidth]{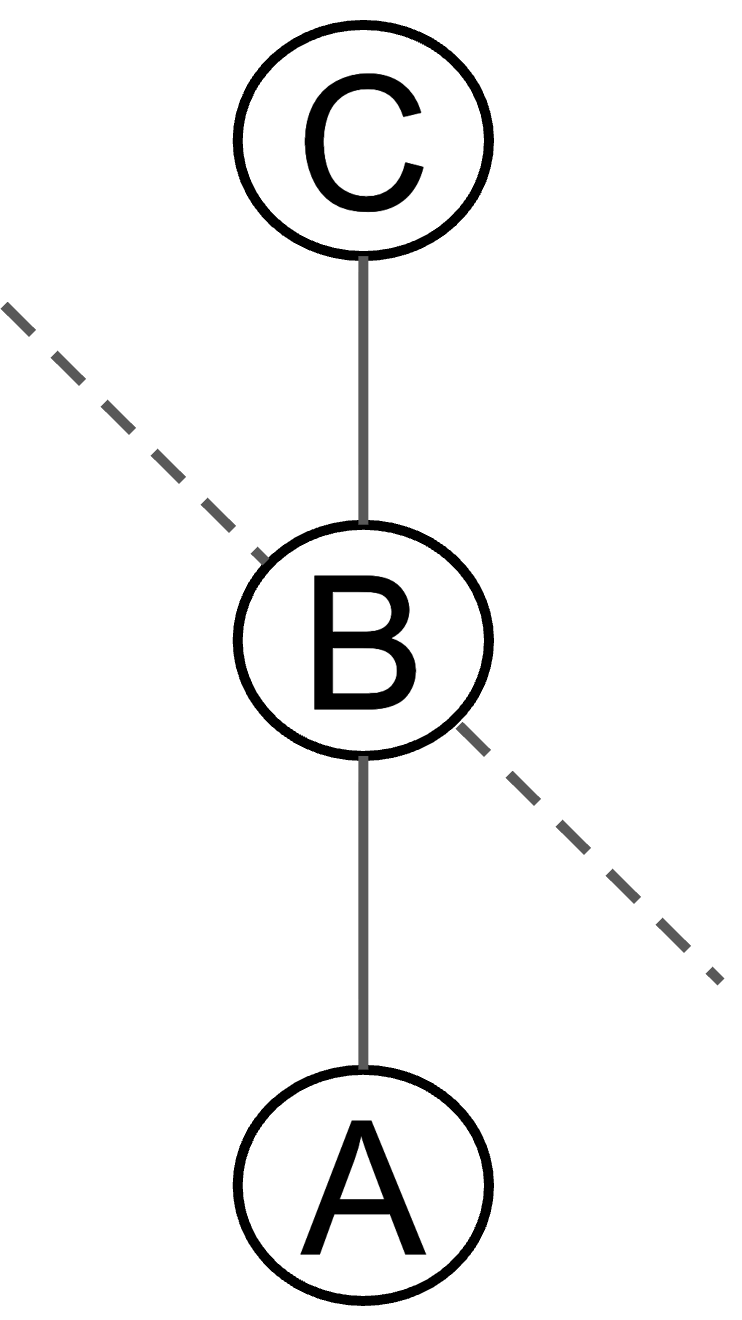}
        \caption{}
        \label{fig: failed spatial}
    \end{subfigure}%
    \hfill
    \begin{subfigure}[t]{0.25\columnwidth}
        \centering
        \includegraphics[width=0.85\columnwidth]{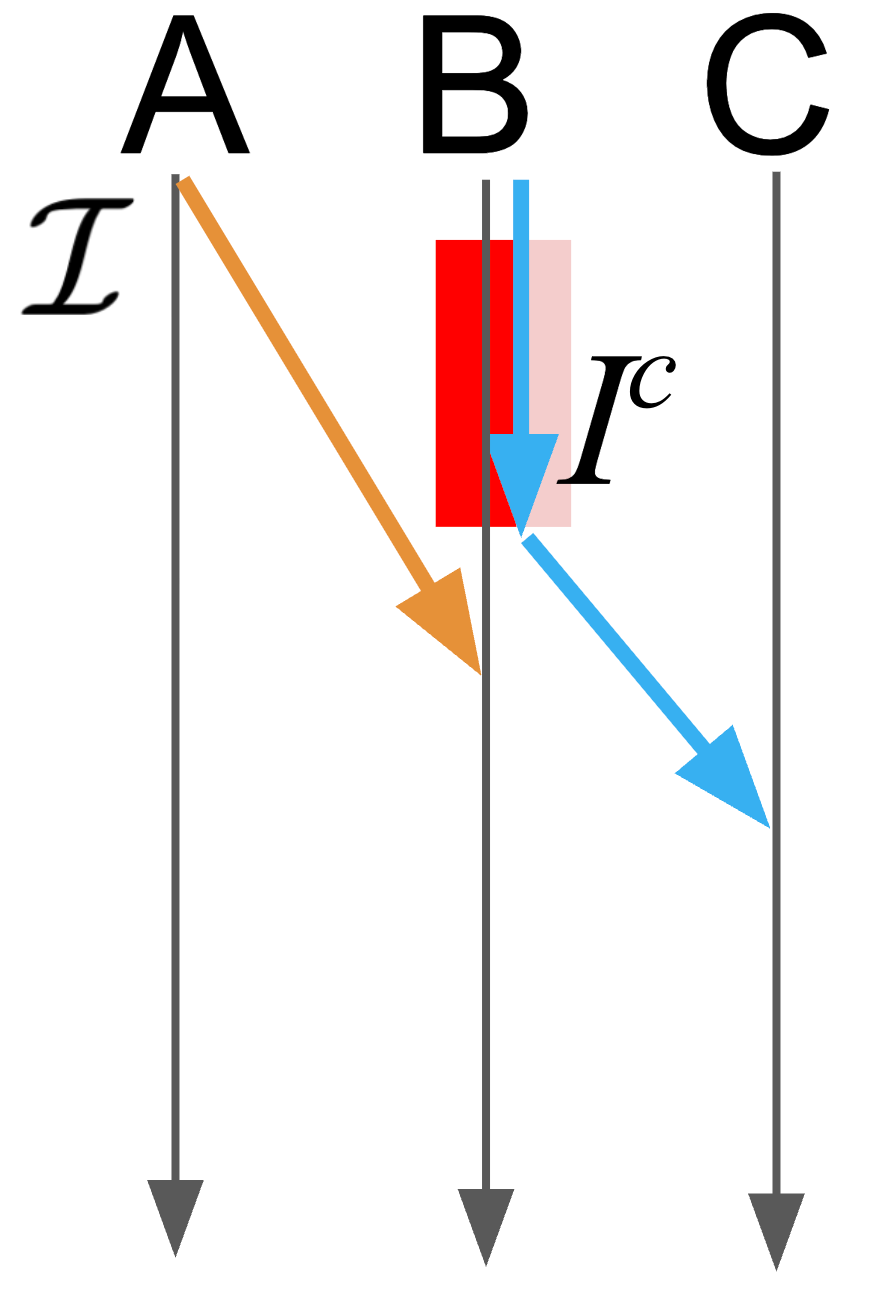}
        \caption{}
        \label{fig: failed time move}
    \end{subfigure}
    \hfill
    \begin{subfigure}[t]{0.24\columnwidth}
        \centering
        \includegraphics[width=0.822\columnwidth]{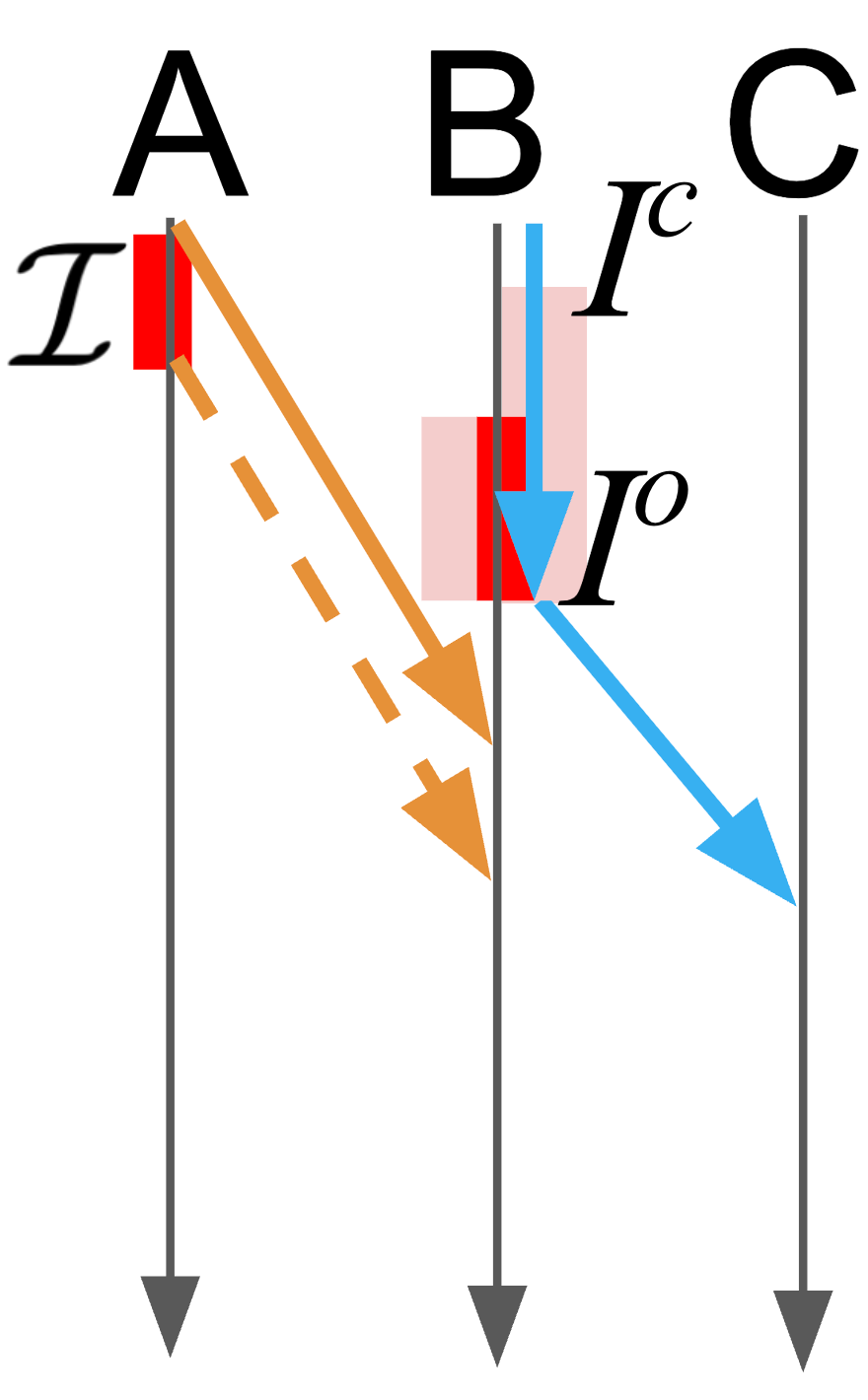}
        \caption{}
        \label{fig: failed time w1}
    \end{subfigure}
    \hfill
    \begin{subfigure}[t]{0.25\columnwidth}
        \centering
        \includegraphics[width=0.875\columnwidth]{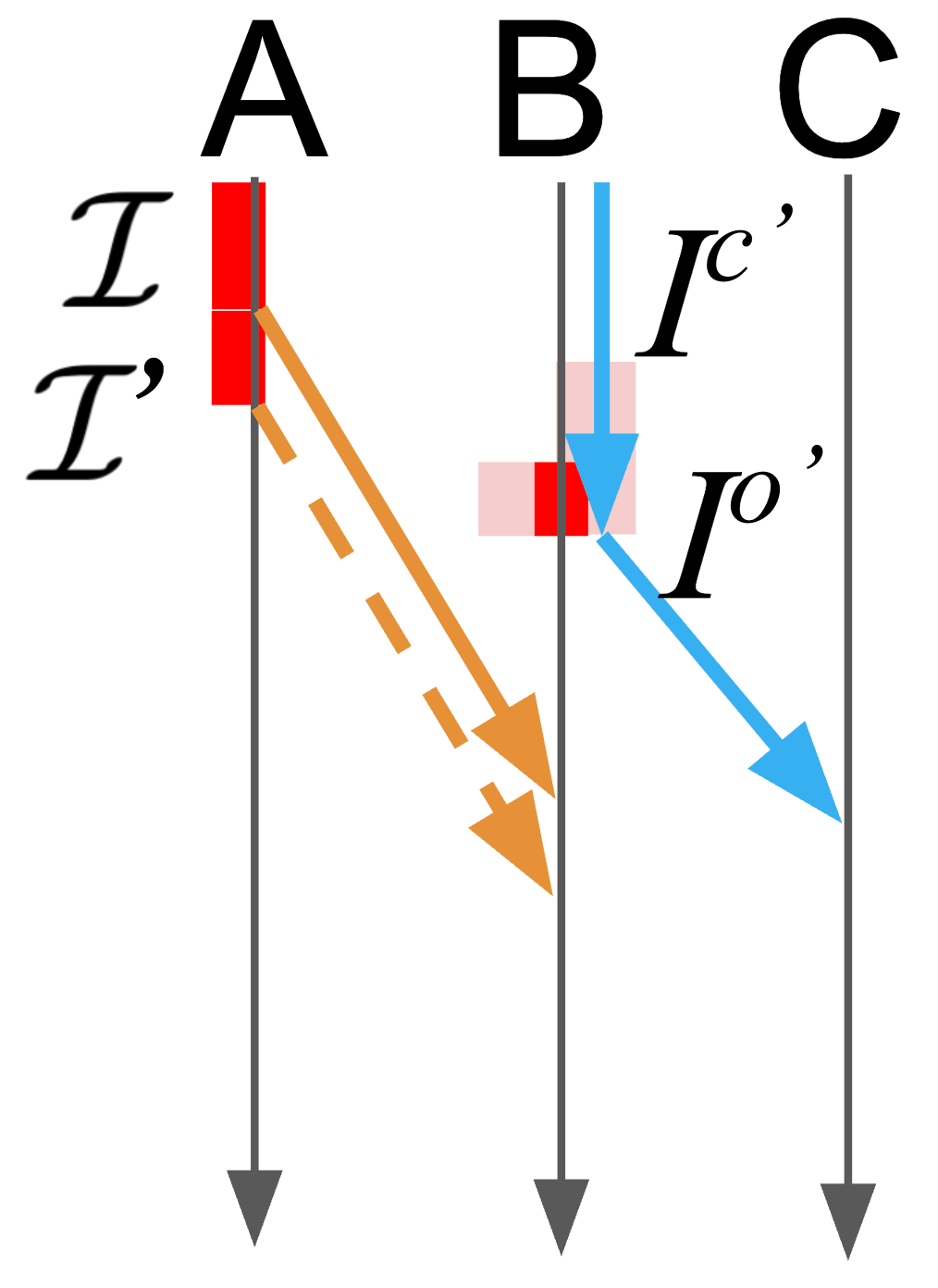}
        \caption{}
        \label{fig: failed time w2}
    \end{subfigure}
    \caption{
        Agent $a_1$ moves from $A$ to $B$, $m_1= \langle (A,B),d_{AB} \rangle@t_1$, and agent $a_2$ waits at $B$ for a while, $w_2= \langle (B,B), d_w \rangle @ t_2-d_w $.
        Then departs to $C$, $m_2=\langle (B,C), d_{BC} \rangle @ t_2 $, where $ t_1 < t_2 < t_1 + d{AB}$.
        The delay could caused by $a_2$'s other constraint. (a) shows the spatial figure.
        A collision is detected between $m_1$ and $w_2$.
        Relative collision intervals are labelled as pink region in (b), (c), and (d).
    }
    \label{fig: shift constraint example}
\end{figure}

Here we attempt to illustrate the problem of shift constraints by at an example of a collision.
Let's say agent $a_1$ is performing $ m_1@\tau_1 $ where $m_1= \langle (A,B),d_{AB} \rangle$.
It collides with $a_2$ at its wait at $B$ $ w_2@(\tau_2-d_w) $, where $w_2= \langle (B,B), d_w \rangle$, as demonstrated in Fig \ref{fig: shift constraint example}.
In this example, $a_2$ departs from $B$ to $C$ before $A$ arrives at $B$, and the collision is due to the size of agents.
There are two possible situations based on the size of the shifted interval $\mathcal{I}$.

First we analyse when $|\mathcal{I}| = 0$, the motion constraint generated for $a_1$ would be $c_1=\overline{\langle a_1, (A,B), [\tau_1,\tau_1]  \rangle}$.
The low-level planner would need to produce a feasible path with least cost while satisfying the constraint $c_1$, however, this is impossible:
there is no immediate successor to a real number,
by definition, as there always exists a number between any two real numbers.
Thus, if $a_1$ performs $m_1@\tau_3$ where $t_1<\tau_3$, there will always exists a number $\tau_3'$, where $\tau_1<\tau_3'<\tau_3$ which yields at a lower cost if $a_1$ performs $m_1@\tau_3'$.
Therefore, a feasible path that satisfies the constraint with \emph{least} cost is mathematically impossible.
Or more generally:
\begin{theorem}
    A least-cost feasible path satisfying the motion or vertex-range constraint does not exist when the constrained interval is closed at its upper bound.
\end{theorem}

Next we analyse the case when $|\mathcal{I}| > 0$.
The motion constraint generated for $a_1$ would be $c_1 = \overline{\langle a_1,(A,B),[\tau_1,\tau_1+\alpha |I^c|) \rangle}$ where $[\tau_1,\tau_1+\alpha |I^c|) = \mathcal{I}$, as shown in Fig \ref{fig: failed time w1}.
Here $I^c$ is the collision interval, which is divided into two portions, $\alpha |I^c|=|\mathcal{I}|$ and $(1-\alpha) |I^c|=|I^o|$, where $\alpha \in (0,1)$, and $I^o$ is the overlapping interval.
Therefore, the vertex range constraint generated for $a_2$ is $c_2 = \overline{\langle a_2,(B,B), I^o \rangle}$.
When we replan the moving agent $a_1$, the move motion which satisfies the new constraint is $m_1@\tilde{t}_e$ where $\tilde{t}_e$ is $\tau_1+\alpha |I^c|$, the ending time of interval $\mathcal{I}$.
This move motion will have a new collision interval ${I^c}'$ with $w_2@\tau_2-d_w$, where $|{I^c}'|=(1-\alpha) |I^c|$.
The motion constraint would be $c_1' = \overline{\langle a_1,(A,B),\mathcal{I}' \rangle}$, where $\mathcal{I}'=[\tilde{t}_e, \tilde{t}_e + \alpha {I^c}')$,
which would expand to $\mathcal{I}'= [\tilde{t}_e, \tilde{t}_e + \alpha (1-\alpha) |I^c|)$.
Following a similar calculation, the next expansion for the moving agent would generate a motion constraint of $c_1' = \overline{\langle a_1,(A,B),[\tilde{t}_e', \tilde{t}_e' + \alpha (1-\alpha)^2 |I^c| ) \rangle}$.
We can see a clear pattern here: every additional constraint added is reduced in size following a geometric series.
This sequence of operations will fall into an infinite loop and will never terminate.

Theorem~\ref{thm:shifting-failure} and Property~\ref{lemma:max_constraint_range} show that patching the termination failure issue in CCBS is non-trivial, and even hints at fundamental limitations of relying on pairs of motion and vertex range constraints to achieve completeness. Such results underscore the necessity of exploring an entirely different approach to tackle the incompleteness challenge in CCBS, or to solve the \mapfr problem optimally in an entirely different way.

%% file: sections/discussion.tex
\section{Discussion}
\subsection{\mapfr with Discrete Time}

Upon its introduction, the \mapfr problem was defined to support arbitrary-duration actions\footnote{Action can be seen as a motion (move or wait) with a specific cost.}~\cite{Thayne2018eICTS}.
However, the considered algorithms (optimal ICTS and bounded sub-optimal $\epsilon$-ICTS and $\omega$-ICTS) restrict wait action to a single unit-time duration.
Subsequent works, such as \cite{Walker2020byclique,Walker2021cbics}, consider optimal solving and adhere to the original \mapfr definition but continue to adopt wait actions with unit durations.
Other works (e.g., \cite{Pavel2021SMTCBS,walker2024byclique}) also adhere to the original \mapfr definition but without introducing unit wait-time restrictions.
This has naturally caused some confusion in the community.
\citet{ccbs} attempted to clarify these differences by presenting a comparative table of existing \mapfr solvers.
This work acknowledges the existence of competing models for \mapfr but the relationship between them is not discussed.
In this section, we attempt to clarify the situation.

We refer to the unit wait-duration variant of \mapfr as \mapfr-Discrete Time (\mapfrdt).
All wait durations are a single unit-time interval under the model of \mapfrdt.
We show that the state-space of \mapfrdt remains countable, unlike \mapfr.
We also show that \mapfrdt is optimally solvable~\cite{Thayne2018eICTS}, also including by CCBS.

\label{Section: completeness of CCBS}
\begin{lemma}
    \label{lemma:MAPFGcountable}
     The Search Space of \mapfrdt is countable, i.e. given a time bound, the number of solutions (feasible and infeasible) for \mapfrdt is countable.
\end{lemma}

\begin{proof}
    For every \mapfrdt instance, an agent's motion at each step can either be move or wait. Let $G = (V,E)$ be the underlying graph. Then, for each agent:
    \begin{itemize}
        \item $|E|$ corresponds to the number of possible move actions between edges.
        \item $|V|$ corresponds to the number of possible wait actions on vertices.
    \end{itemize}
    Hence, the total number of distinct motion actions for an agent is finite, $|E|+|V|$ for any \mapfrdt instance.
    Each agent's path is a concatenation of motions, which can be represented as $[E \cup V]*$, following regular expression syntax, where:
    \begin{itemize}
        \item $E=\{ e_1, e_2,...e_m \}$ denotes the set of all move motions in the graph; and
        \item $V=\{ v_1, v_2,...v_n \}$ denotes the set of all wait motions in the graph.
    \end{itemize}
  Since any regular expression is countable, each agent's possible path is also countable.
  Furthermore, each motion has an associated cost comprised of the following:
  \begin{itemize}
      \item Each move action $e_j$ has a fixed cost $e_j.D$, and;
      \item The wait action, $w_v$, has a fixed cost, typically 1 unit cost per wait.
  \end{itemize}
Therefore, the total cost of a path is a finite sum of fixed-cost components:
$\sum_{j=1}^{m} n_j \cdot e_j.D+n_w \cdot w$ where $n_j$ is the number of times edge $e_j$
appeared in the path, and $n_w$ is the number of wait actions. Since all
constants are fixed and the multipliers range over natural numbers, the set of
possible path costs is countable. The full search space is a cross-product of
all agents' possible paths, which is the product of a finite number of countable
sets, since the number of agents is finite for any \mapfr problem. Thus, the
search space for any \mapfrdt instance is countable.
\end{proof}

\begin{theorem}
    CCBS is sound with motion constraint, solution complete, and guaranteed to return an optimal solution for the \mapfrdt problem, if one exists.
\end{theorem}

\begin{proof}
The two conditions required for the correctness of the CCBS framework in  Lemma~\ref{lemma:condition} hold in the context of \mapfrdt. Furthermore, from Lemma~\ref{lemma:MAPFGcountable}, we know that the solution space for any \mapfrdt instance is countable.
Since all motion costs are positive and fixed, and the number of agents is finite, this implies that the number of valid (collision-free) solutions under any given cost bound is also finite. Therefore, if a solution exists, there exists a finite subset of the search space bounded by the cost of the optimal collision-free solution.
CCBS resolves collisions incrementally by adding pairwise constraints that eliminate at least one colliding solution in each iteration. Because the number of such candidate solutions is finite, the algorithm must eventually converge to a collision-free solution if one exists. \textit{Thus, if a solution exists, CCBS is solution complete, i.e. it will always find a solution if one exists.} Moreover, CCBS explores the solution space in a best-first manner based on path cost. This ensures that the first collision-free solution it encounters is the one with the lowest possible cost. \textit{Thus, CCBS always returns the least-cost valid solution, if one exists.}
Finally since CCBS is sound (for \mapfrdt), every returned solution is verified
to be both collision-free and adherent to all imposed constraints.
\end{proof}

We have demonstrated that the \mapfrdt problem can be effectively solved using
enumerative algorithms such as CBS. However, as per Theorem \ref{theorem:1},
applying such enumerative methods to the original \mapfr model leads to
infinite expansions due to its continuous temporal domain. This raises an
important open question: \textit{how can we solve the \mapfr problem
efficiently, or even determine whether it is solvable in the general case?}

\subsection{Toward Solvability for \mapfr}

It is unclear whether theoretical guarantees which hold for \mapfrdt also hold for \mapfr.
This is due to several generalisations introduced in \mapfr, including continuous temporal reasoning and more complex spatial interactions between agents.
As a result, standard approaches for verifying solvability, such as running a polynomial-time solver prior to invoking CBS, are no longer applicable.

There is currently no complete solver for \mapfr.
Several suboptimal solvers have been proposed \cite{park2023subMAPFR1,ijcai2024subMAPFR2}; however, these either lack completeness guarantees or rely on the completeness of CCBS, which—as discussed—applies only to the \mapfrdt variant.
Consequently, we currently do not have an efficient suboptimal method for verifying the solvability of a \mapfr instance.

\begin{figure}[h]
    \centering
     \begin{subfigure}[t]{0.25\columnwidth}
        \centering
        \includegraphics[width=1\columnwidth]{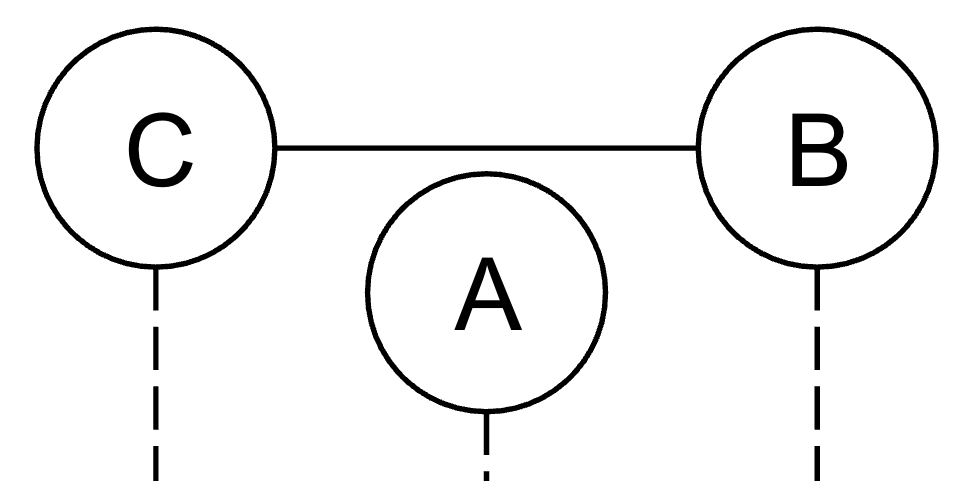}
        \caption{An agent waiting at vertex $A$ also blocks edge $\langle B,C\rangle$}
        \label{fig: wait agent block edge}
    \end{subfigure}%
    \hfill
    \begin{subfigure}[t]{0.33\columnwidth}
        \centering
        \includegraphics[width=1\columnwidth]{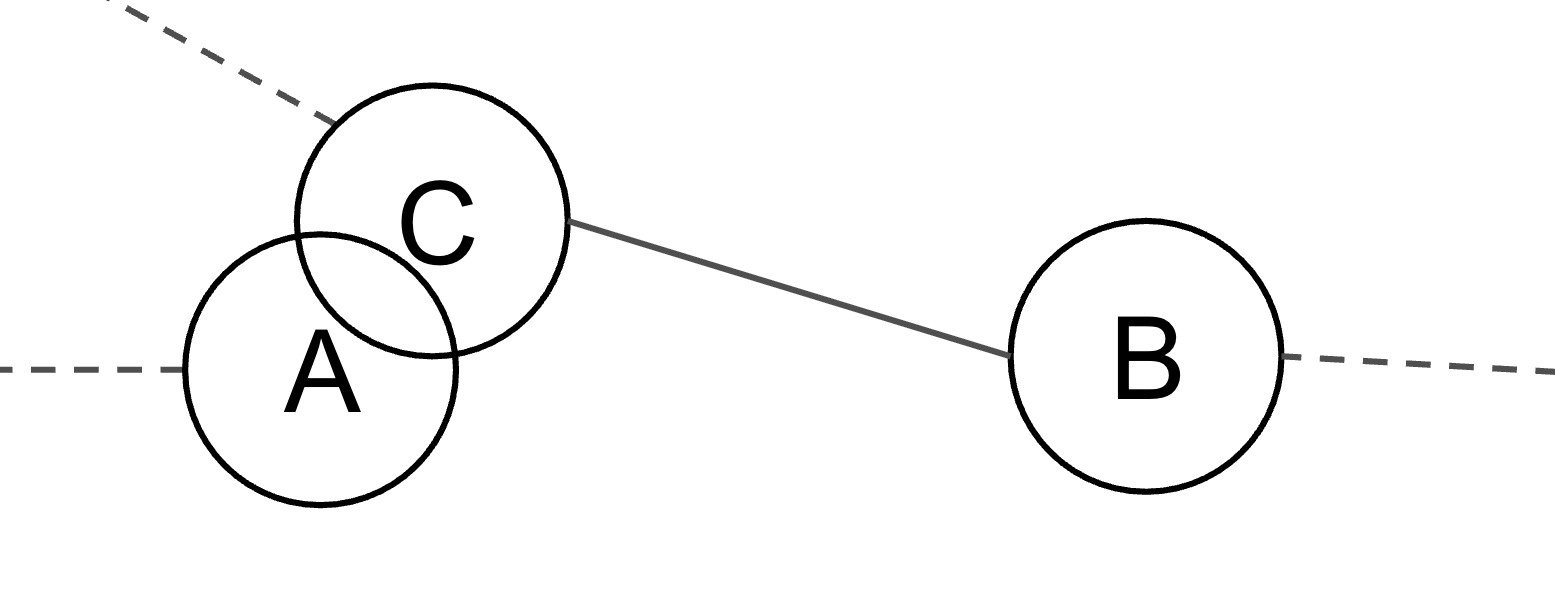}
        \caption{An agent waiting at vertex $A$ also blocks vertex $C$}
        \label{fig: wait agent block vertex}
    \end{subfigure}
    \begin{subfigure}[t]{0.40\columnwidth}
        \centering
        \includegraphics[width=1\columnwidth]{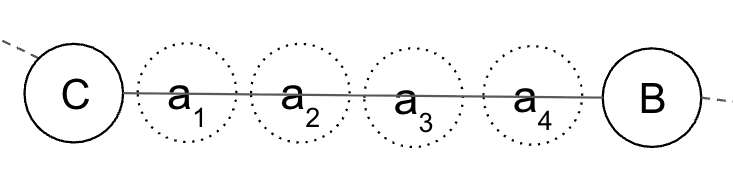}
        \caption{Multiple agent moving along $\langle B,C \rangle$}
        \label{fig: wait agent buffering}
    \end{subfigure}
\caption{Example of a waiting agent interfering with non-adjacent edges or vertices}
\label{fig: wait agent blocks}
\end{figure}

As we have demonstrated, the primary challenge in solving \mapfr lies in handling continuous wait durations.
One might attempt to sidestep this issue using a reduction to the pebble motion problem \cite{kornhauser1984pebble}.
In the pebble motion problem, one agent/pebble moves at one time, which substantially simplifies coordination allowing classical MAPF solutions to be computed in polynomial time \cite{roger2012MAPFsolved}, with $O(N^3)$.
However, \mapfr introduces additional complexity not present in classical MAPF.
First, agents are modelled with physical dimensions  (e.g., circular agents with a given radius).
This physical embodiment fundamentally changes moving and waiting behaviour in the environment.
In particular, a waiting agent can physically overlap and obstruct nearby paths, depending on its position and size (see Figure \ref{fig: wait agent blocks}).
This spatial interference significantly complicates conflict resolution and motion planning.
Second, in \mapfr multiple agents can traverse a long edge at the same time.
This characteristic introduces a \emph{buffering} effect, where a number of occupied vertices are freed up temporarily, allowing movements that would not be possible otherwise (see Figure \ref{fig: wait agent blocks}).
Thus, there exist instances which are solvable in \mapfr, but for which there exists no corresponding pebble motion solution.
We define three types of spatial features that are unique to \mapfr:
\begin{itemize}
    \item \textit{Edge Overlapping}: This occurs when a waiting agent blocks the traversal of an edge due to its physical presence. Formally, Let  agents $a_i$ and $a_j$ be such that:
    \begin{itemize}
        \item $a_i$ is waiting at a vertex $v_k$, represented as $w_i=\langle e_k.D, v_k \rangle$; and
        \item $a_j$ is moving on the edge $e_k$, represented as $m_j=\langle e_k.D, e_i \rangle$,
    \end{itemize}

    where $e_k.D$ is the duration to traverse edge $e_k$. We say that vertex $v_k$ is \textit{Edge Overlapping} with edge $e_k$, if $InCollision(w_i@t,m_j@t) = true$ for some time $t$. (Fig. \ref{fig: wait agent block edge}).

    \item \textit{Vertex Overlapping}: This occurs when the physical sizes of two waiting agents cause them to collide, even if they are located at distinct vertices. Let agents  $a_i$ and $a_j$ be such that:
    \begin{itemize}
        \item $a_i$ is waiting at a vertex $v_{k1}$, represented as $w_i=\langle I, v_{k1} \rangle$; and
        \item $a_j$ is waiting at another vertex $v_{k2}$, represented as $w_j=\langle I, v_{k2} \rangle$,
    \end{itemize}
    for some non-zero duration $I$. We say that vertex $v_{k1}$ is \textit{Vertex Overlapping} with vertex $v_{k2}$, if \\ $InCollision(w_i@t,w_j@t) = true$ for some time $t$ (Fig. \ref{fig: wait agent block vertex}).
    \item \textit{Buffering}: This occurs when multiple agents can traverse on the same edge at the same time without colliding to each other.
    We say the edge $e_i$ is buffering, when a set of agents $A'$ are executing $\langle e_i,|e_i| \rangle$ at the same time instance $t$ and no collision exists between any two agents in $A'$ at time $t$.
\end{itemize}
Thus, the question of whether \mapfr is solvable at all remains unanswered.

\subsection{Essential Conditions for Establishing Completeness and Optimality}

Mapping the MAPF problem to a best-first tree search framework is one of the most popular approaches in the field.
However, without careful problem modelling, search algorithms risk losing their guarantees of completeness and optimality.
Tree search algorithm provides completeness and optimality guarantee by pushing the lower bound enumeratively until a goal is reached.
This process requires satisfying two fundamental conditions:
\begin{enumerate}
    \item The state space must be finite, given any upper bound on the solution cost.
    \item All step costs must be non-negative.
\end{enumerate}

Following the CBS structure, for a given CT node $N$, the state is defined as the set of all solutions that satisfy $N.constraints$.
Due to the uncountably infinite ways to split a wait motion, each agent has an uncountably infinite number of possible paths that satisfy its constraints, resulting in an uncountable set of possible solutions for each state.
In order to prove the state space is finite for CCBS, two conditions must satisfied:
\begin{enumerate}
    \item There is finite amount of cost value given any upper bound. This can be achieved by proving a finite amount of cost value will dominate all other real-valued cost.
    Or proving there is a minimum step size between any two cost value.
    \item There is a finite number of possible states for each possible cost value.
        The state definition is tightly connected with the constraint definition; thus, this can be achieved by a type of constraint that partitions the solution space into a finite number of states.
\end{enumerate}

Generally speaking, to establish completeness guarantees for CCBS, requires a novel type of constraint that not only excludes subsets of colliding solutions but also ensures that a finite number of such constraints can raise the solution cost to any arbitrary finite value $C^*$.

\subsection{Numerical issue with \mapfr Solvers}
Supposing that a complete and optimal solver is possible,
additional challenges arise when attempting to implement these theoretical algorithms. For practical use, computed plans may not be feasible. Meanwhile, for scientific studies empirical claims may not be reproducible,
complicating comparisons between competing algorithms and undermining theoretical claims.
Both challenges arise because modern computers cannot represent real numbers exactly due to hardware limitations.
One of the most widely used representations is the IEEE-754 floating-point format, which stores approximations of real values rather than exact quantities.
Such approximations introduce precision loss, an issue that has been extensively studied in fields such as computational geometry, where various techniques have been proposed to mitigate numerical error \cite{Priest1991multi_term,bailey1993multi_digit}.
As a result of approximation, floating-point representations can also produce inconsistent results from mathematically equivalent computations under certain circumstances, such as double rounding.
Tree-search–based algorithms such as CCBS are particularly vulnerable to these inconsistencies, often to a greater extent than to raw precision loss alone.
Even a minor perturbation in a node’s cost can redirect the search toward a different branch, ultimately yielding a different solution.
These inconsistencies draw into question not only comparisons between competing algorithms but also between different implementations of the same algorithm.
Even in a single implementation, processing the same conflict on different branches can lead to distinct and non-equivalent constraints.

\begin{figure}[t]
        \centering
        \includegraphics[width=0.3\columnwidth]{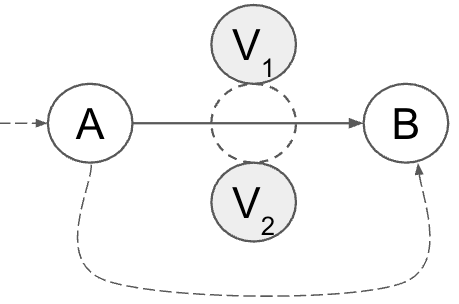}
        \caption{Agent $a_1$ is moving from $A$ to $B$, and $a_2$ and $a_3$ are stationary at $V_1$ and $V_2$.
        This is only a portion of the graph; dashed edges represent how this subgraph interacts with the remainder of the graph.
        The dashed circle shows that the gap between $V_1$ and $V_2$ is just enough for one agent to traverse through.}
\label{fig: FP_problem}
\end{figure}

Figure \ref{fig: FP_problem} illustrates an example of unbounded cost inflation caused by inconsistencies in floating-point computation. In this scenario, agent $a_1$ attempts to traverse edge $(A,B)$, while agents $a_2$ and $a_3$ are stationary at $V_1$ and $V_2$, respectively. The separation between $V_1$ and $V_2$ is exactly $2r$, which permits the passage of a single agent without collision.
Due to numerical precision loss, the validator may incorrectly report the traversal of $(A,B)$ as infeasible.
A more problematic effect arises: if the move $\langle(A,B), d_{AB}\rangle@t_1$ is executed at time $t_1$, it may be evaluated as collision-free with respect to both $a_2$ and $a_3$.
However, within the same execution of the solver, delaying the same move to a later time $t_2 > t_1$ may lead the validator to report a collision with either $a_2$ or $a_3$.

Under floating-point representation, this inconsistency may stem from several sources.
Precision generally decreases as the magnitude of the number increases, so representing $t_2$ may incur greater rounding error than $t_1$.
Additional factors include differences in operation ordering or double rounding effects, even under deterministic implementations where evaluation order may still vary across expressions.
For example, if $checkCollision(a_i, a_j)$ determines collision by enumerating all pairwise motions, the feasibility outcome may differ depending on argument ordering (i.e., $a_i = a_1$ versus $a_j = a_1$), which may change at different iterations.
This is because floating-point numbers do not guarantee many fundamental arithmetic properties, such as associativity.
Alternatively, if the traversal from $A$ to $B$ is deemed infeasible, the only remaining detour to reach $B$ may incur arbitrarily large cost, thereby inflating the computed “optimal” solution cost without bound.
Such inconsistencies can arise at multiple levels of the solver, including:
\begin{enumerate}
    \item {\bf Validator}: Collision checking can be done analytically, such as solving a quadratic equation under the CCBS kinematics assumptions, constant velocity and circular agents.
    It can also be done numerically, such as binary search.
    Both methods are approximations, but slightly different.
    They may report differently when checking the collision over the same pair of motions.
    This may cause the solution returned from one solver to be deemed infeasible by another solver, or the false positive collision may also introduce unnecessary cost increase in the solution.
    \item {\textbf{Numerical representation:}} Floating point is the default choice, but alternatives such as exact arithmetic\cite{valerie2005BadicNumber}, interval arithmetic\cite{intervalMath}, and other approximation representations\cite{Priest1991multi_term,bailey1993multi_digit} also exist. Different numeric formats, or even the same format at different precision levels, can lead to disagreement on the correctness or optimality of solutions.
    \item {\bf Tolerance values}: a commonly used method to check equality is use a value $\epsilon$, i.e. two values are deemed as equal when their difference is less than $\epsilon$, $a=b$ if $|a-b|<\epsilon$. However, different implementations may choose different values of $\epsilon$. Even with same $\epsilon$ values, results can differ because of the ordering of operands, and the specific implementation of the comparison operator.
    \item {\bf Compiler and hardware optimisation}: different optimisation techniques may also result in different results to the same calculation, including the usage of fused multiply-add (FMA), handling method of denormal number (FTZ and DAZ).
\end{enumerate}

Reliable measurement of progress is essential for advancing research.
How implementations can be made to converge on a common notion of optimality, or whether this is achievable at all, remains an open question.
We argue that the immediate priority should be agreement on feasibility prior to optimality, which can be enforced through a shared validator implementation.
This approach directly addresses numerical inconsistencies (1)-(3). With careful experimental controls (e.g., similar software and hardware selection), we believe this approach can also mitigate numerical inconsistency (4). In practice, solvers should be treated as suboptimal, avoiding definitive empirical claims and reporting both solution cost and runtime performance.

%% file: sections/conclusion.tex
\section{Conclusion}
In this paper, we examined the limitations of the CCBS algorithm, which is not only a cornerstone method for \mapfr but also the only optimal \mapfr solver.
First, we showed that the CCBS model for waiting motions is ambiguous and gives rise to two possible interpretations for how to handle the corresponding conflicts: {\em motion constraints} and {\em vertex range constraints}.
We analyse both possibilities
and prove that they are unsound and/or incomplete.
In a subsequent result, we generalised the vertex range constraints into shift constraints, and examined the limitations of the shift constraint to show that this class of constraints may suffer from termination failure.
These findings fundamentally undermine the completeness and optimality
guarantees of CCBS and its many enhancements and derivatives. For instance,
SMT-CCBS, discussed in the latter part of \cite{ccbs}, tackles the \mapfr
problem using Satisfiability Modulo Theories. While this may appear to be a
distinct method, its incorporation of constraints into propositional logic
follows the same formulation as CCBS. As a result SMT-CBS inherits the same set
of problems:  unsound termination or unbounded expansion and eventual failure.
All improvements directly built upon CCBS, such as Bypass, Biclique Constraints \cite{walker2024clique}, and Disjoint Splitting \cite{improve-ccbs} are similarly affected.

To clarify the theoretical foundations of \mapfr, we surveyed current state of the literature and distinguished \mapfrdt as a restricted variant of general \mapfr in which only unit-cost waiting durations are allowed.
This problem can be solved optimally by $\epsilon$-ICTS and by CCBS.
In contrast, the solvability of general \mapfr remains an open question.
To understand this gap, we examined the problem from two complementary perspectives. From the standpoint of completeness, we identified the additional challenges that arise when extending classical MAPF to obtain a complete solver for \mapfr.
From the standpoint of optimality, we characterised the conditions necessary to guarantee optimality for tree-search–based \mapfr solvers, including CCBS.
Finally, we highlighted practical implementation challenges, particularly those related to numerical representation.

Looking ahead, we believe that substantial progress in this area will require either the introduction of new constraint mechanisms within the CCBS framework or the development of fundamentally new algorithmic approaches.
From a theoretical perspective, a key open question is whether \mapfr is solvable and, in particular, whether it can be effectively addressed using CBS-style conflict-based tree search.
From a practical perspective, the field would also benefit from standardized and reproducible evaluation methodologies, enabling meaningful comparisons between solvers and facilitating measurable cumulative progress.

\section{Acknowledgements}
This research was funded in part by a generous gift from Amazon Robotics.

%% file: sections/Appendix.tex
\section{Computing Collision Interval}
\label{App: collision interval}
\begin{figure}
    \centering
    \includegraphics[width=0.5\columnwidth]{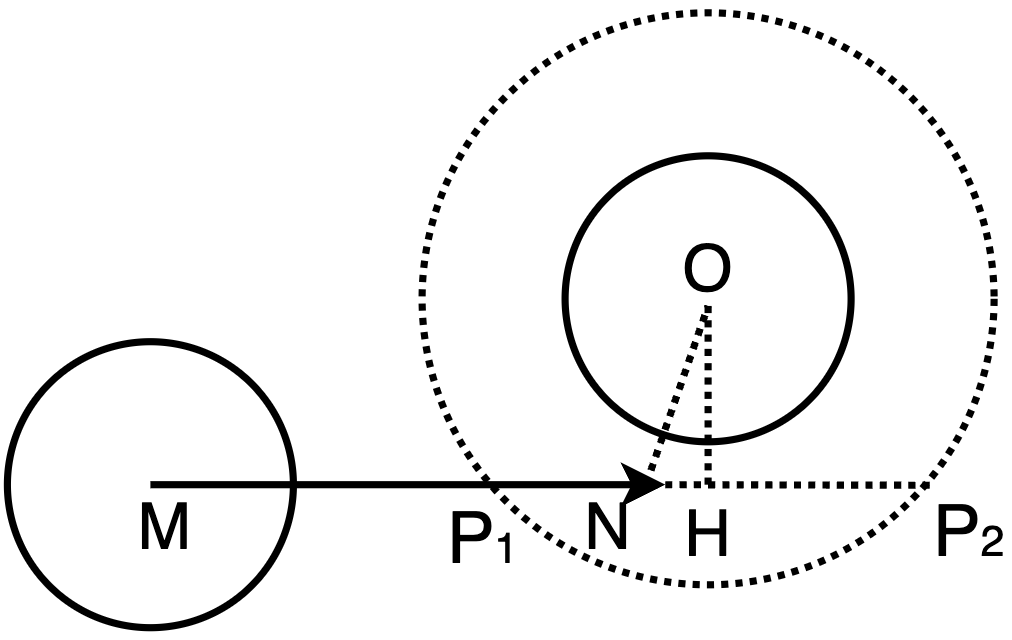}
    \caption{The moving action is $ M@t_0 \rightarrow N@t_1 $, and the waiting agent is parking at vertex $O$. The radius of the dashed circle shown in the figure is $2*r$.
    $P_1$ is the point where two agents start to collide, and $P_2$ is the point the moving agent leaves the collision region if it keeps moving along the extension of $(M, N)$ segment.
    $(O,H)$ is the perpendicular bisector of $(P_1, P_2)$.
}
\label{fig:wait_exp}
\end{figure}

In this section we give a detailed example on how collision interval is computed in implementation.
Following the conceptual example in Figure \ref{fig:wait_exp}, if agent $a_i$ waits at a vertex $O$ infinitely with wait action $w_i@t$, and another agent $a_j$ moves from vertex $M$ to $N$ with action $m@t_0$, $t_0 \gg t$.
The \emph{collision interval} between $w@t$ and $m@t_0$, also the time range of the range constraint for $a_i$, is $(t_0+||(M,H)||_2-||(P_1,H)||_2,t_1)$, $t_0$ is the moving action start time, $P_1$ is the location point on edge $(M,N)$ where $a_j$ starts overlapping with $a_i$ if its centre is on $P_1$, and $t_1$ is the moving action end time.

\section{Counter Example}
\label{App: Counter Example}
\begin{figure}
    \centering
    \includegraphics[width=0.5\columnwidth]{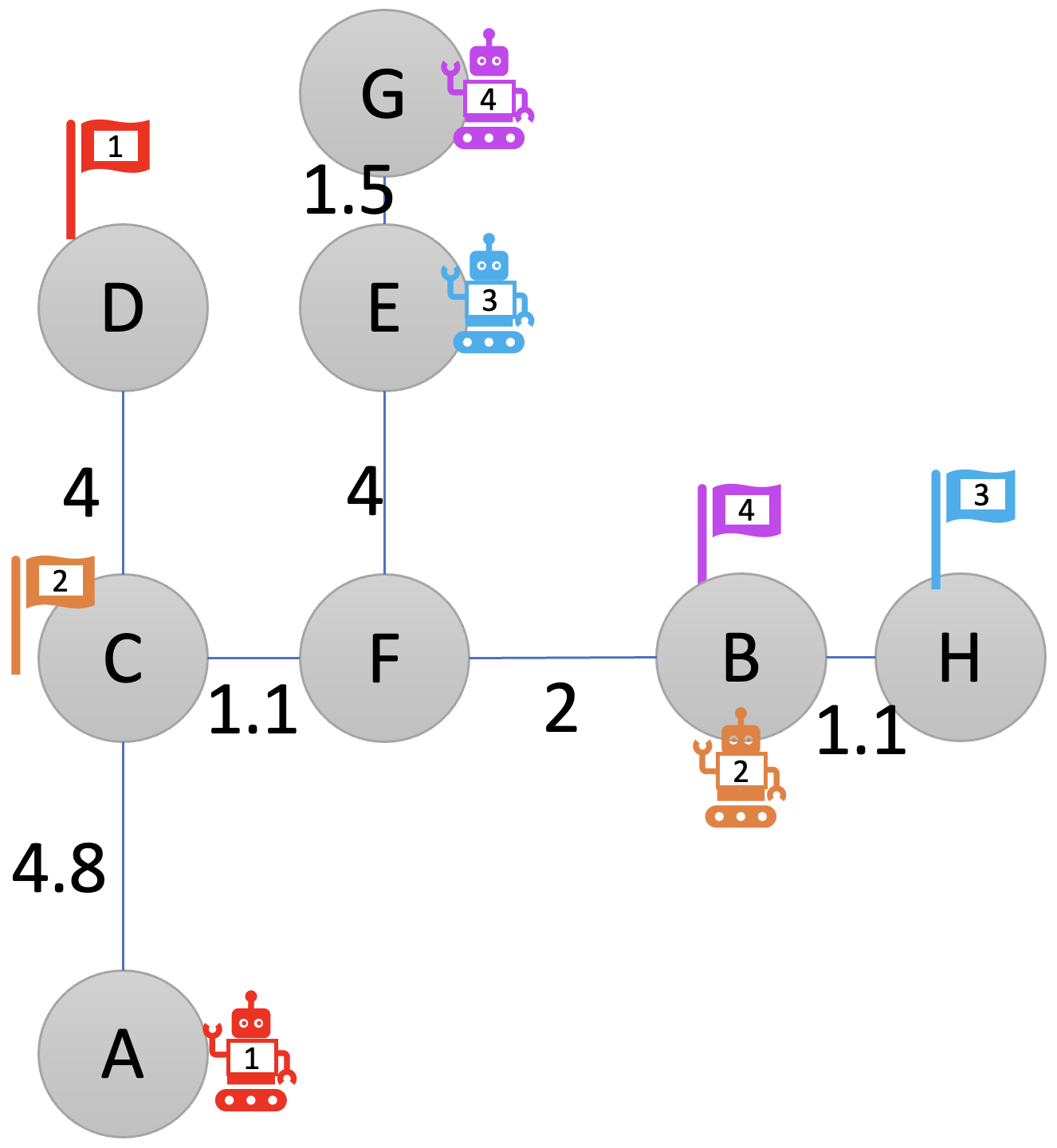}
\caption{Agent $a_1$(red) is trying to move from $A$ to $D$, agent $a_2$(orange) is trying to move from $B$ to $C$, agent $a_3$(blue) is trying to move from $E$ to $H$, and agent $a_4$(magenta) is trying to move from $G$ to $B$. The cost of traversal is labelled on each edge. All agents have a radius of 0.5. }
\label{fig:counter imp}
\end{figure}

We now present another example where eliminating feasible solution leads to a suboptimal solution (Fig\ref{fig:counter imp}).
A plan for an agent can be denoted by chaining these notations together.

In this example agent $a_1$ moves from $A$ to $D$, agent $a_2$ moves from $B$ to $C$, agent $a_3$ moves from $E$ to $H$, and agent $a_4$ moves from $G$ to $B$.
The first conflict arises between agents $a_1$ and $a_2$ on $m_1=\langle (A,C), 4.9 \rangle @ 0$ and $w_2=\langle (C,C),\infty \rangle @ 3.1$.
The constraints generated by the CCBS implementation are $\overline{\langle a_1, m_1, [0,\infty)  \rangle}$ and $\overline{\langle a_2, C, [3.9,4.9)  \rangle}$.
They eliminate the solution of agents $a_1$ and $a_2$ performing wait motions prior to departing.
This solution allows agents $a_3$ and $a_4$ to move first, which avoids a collision.

The final CCBS solution has a cost of 34.6 (costs are rounded to 1 d.p.):\\
$a_1$: A@0 $\rightarrow$ C@4.8 $\rightarrow$ D@8.8\\
$a_2$: B@0 $\rightarrow$ B@3.2	$\rightarrow$  F@5.2 $\rightarrow$ F@5.2 $\rightarrow$ C@6.3\\
$a_3$: \textbf{E@0} $\boldsymbol{\rightarrow}$ \textbf{E@2.6} $\rightarrow$ F@6.6 $\rightarrow$ B@8.6 $\rightarrow$ H@9.7\\
$a_4$: \textbf{G@0} $\boldsymbol{\rightarrow}$ \textbf{G@2.1} $\rightarrow$ \textbf{E@3.6} $\boldsymbol{\rightarrow}$ \textbf{E@4.0} $\rightarrow$ \\
F@8.0 $\rightarrow$ B@10.0

Here we provide a handcrafted solution with a much smaller cost of 32.1. \\
$a_1$: \textbf{A@0} $\boldsymbol{\rightarrow}$ \textbf{A@0.5}	$\rightarrow$ C@5.3      $\rightarrow$ D@9.3\\
$a_2$: B@0 $\rightarrow$ F@2	$\rightarrow$ C@3.1    $\rightarrow$ \textbf{C@3.9} $\boldsymbol{\rightarrow}$ \textbf{F@5} $\boldsymbol{\rightarrow}$ \textbf{F@5.6} $\boldsymbol{\rightarrow}$ \textbf{C@6.7}\\
$a_3$: E@0 $\rightarrow$ F@4	$\rightarrow$ B@6      $\rightarrow$ H@7.1\\
$a_4$: G@0 $\rightarrow$ E@1.5	$\rightarrow$ E@3      $\rightarrow$ F@7    $\rightarrow$ B@9

In this solution, both agents $a_1$ and $a_2$ are required to wait so that agents $a_3$ and $a_4$ will not need to.
However, CCBS's range constraint eliminates such a solution, therefore vertex and motion constraint pair violates Claim \ref{clm:sound}, hence they are not sound.
This goes to show that the current CCBS implementation is incomplete and may return suboptimal solutions.
Note that, the CCBS implementation expanded 780 high-level node before finding the solution, thus, this instance is non-trivial, i.e., not solvable by hand.

